\documentclass{article}
\usepackage{spconf,amsmath,graphicx}
\usepackage{multirow}
\usepackage{graphicx}
\usepackage{bm}
\usepackage{lineno}
\usepackage{url}
\usepackage{tikz}
\usepackage{verbatim}
\usetikzlibrary{positioning}
\usepackage{calc}
\usepackage{amsfonts}

\title{FCEM: A Novel Fast Correlation Extract Model For Real Time Steganalysis of VoIP Stream via Multi-head Attention}
\name{Hao Yang, ZhongLiang Yang$^{\star}$, YongJian Bao, Sheng Liu, YongFeng Huang$^{\star}$
\thanks{This work was supported in part by the National
Key Research and Development Program of China under
Grant 2018YFB0804103 and the National Natural Science Foundation of China (No.U1836204, No.U1705261 and
No.61862002)}
\thanks{$^{\star}$Corresponding author}
}
\address {Department of Electronic Engineering, Tsinghua University, Beijing 100084, China}

\begin{document}
%\ninept
%
\maketitle
\begin{abstract}
Extracting correlation features between codes-words with high computational efficiency is crucial to steganalysis of Voice over IP (VoIP)  streams. In this paper, we utilized attention mechanisms, which have recently attracted enormous interests due to their highly parallelizable computation and flexibility in modeling correlation in sequence, to tackle steganalysis problem of Quantization Index Modulation (QIM) based steganography in compressed VoIP stream. We design a light-weight neural network named Fast Correlation Extract Model (FCEM) only based on a variant of attention called multi-head attention to extract correlation features from VoIP frames. Despite its simple form, FCEM outperforms complicated Recurrent Neural Networks (RNNs) and Convolutional Neural Networks
(CNNs) models on both prediction accuracy and time efficiency. It significantly improves the best result in detecting both low embedded rates and short samples recently. Besides, the proposed model accelerates the detection speed as twice as before when the sample length is as short as 0.1s, making it a excellent method for online services. 
\end{abstract}
\section{Introduction}
Steganography tries to hide messages in plain
sight while steganalysis tries to detect their existence secret data from suspicious carriers. VoIP is a real-time service which enables users to make phone calls through IP data networks. It is utilized in many popular apps such as Wechat, Skype and 
Snapchat, which are widely used for instant messaging in daily life.  Compared with traditional steganographic audio 
carriers, VoIP possesses many particular advantages, including instantaneity, a large mass of carrier data, high covert 
bandwidth and flexible conversation length, making it a very popular scheme for covert communication\cite{VOIP-1,VoIP-2,31-Xiao2008An}. However, like many 
other security techniques, VoIP-based steganography might also be employed by lawbreakers, terrorists and hackers for 
illegitimate purposes, which causes serious threats to cybersecurity. Thus, it is crucial to develop a powerful and practiacal steganalysis tool for VoIP stream.

% For example, in image steganalysis, Qian et al. \cite{Qian2015Deep} 
% firstly introduced convolutional neural networks (CNN) in order to detect the existence of steganographic content. The 
% proposed model can capture the complex dependencies that are useful for steganalysis which can acheive state of art result. 
% In text steganalysis, Wen et al. \cite{Wen2019Convolutional} propose a novel text steganalysis model based on convolutional 
% neural network,  which is able to capture complex dependencies and learn feature representations automatically from the 
% texts. The proposed method can effectively detect different kinds of text steganographic algorithms and achieve comparable 
% or superior performance for a wide variety of text size. 
Recent years have witnessed a variety of neural 
network models that boost the performance of audio steganalysis. For example, Paulin $et\ al.$ \cite{Paulin2016Audio} firstly employed deep belief networks to solve speech steganalysis problems. They calculated Mel Frequency  Cepstrum Coefficient (MFCC) and deep belief networks (DBN) served as a classifier. Chen $et\ al.$ \cite{24-Chen2017Cnn} used Convolutional Neural Network to detect LSB steganography in time domain. Lin $et\ al.$ \cite{Lin2018RNN} proposed Codeword Correlation Model (CCM), which used recurrent neural network (RNN) to extract correlation features in speech steganalysis. The authors in \cite{yang-cnnlstm} proposed a model to combine the strength of bi-direactional Long Short-Term Memory (LSTM) and Convolutional Neural Network to conduct audio steganalysis and achieved currently state-of-the-art result.

Obviously, recent works often switch between two types of deep neural network (DNN): RNNs 
and CNNs. RNNs perform better in sequential architecture and can capture long-range dependencies but have low computational efficiency. CNN whose hierarchical structure is good at extracting local or position-invariant features but is hard to capture long-range dependencies. Compared with RNNs and CNNs, the attention mechanism  \cite{att} is more flexible in modeling sequence than RNN/CNN, and is more task/data-driven when modeling dependencies. Unlike sequential models, its computation can be easily and significantly accelerated by existing distributed/parallel computing schemes which are very suitable for real time steganalysis in scenarios like VoIP. However, to the best of our knowledge, steganalysis method solely based on attention to extract correlation features has not been studied for VoIP steganalysis or even for audio steganalysis. Thus, it's very promising to design a model that thoroughly based on attention mechanism to steganalysis.

In this paper, we proposed a light-weight and RNN/CNN-free neural
network named Fast Correlation Extract Model (FCEM) to conduct steganalysis of  QIM steganography \cite{Chen2002Quantization} in VoIP streams. In the proposed model, compressed speech frames are firstly converted to dense vector based on an embedding matrix. Then, a variant of attention called "multi-head attention" is used to capture the correlations or frame context dependency which produce context-aware representations for compressed speech frames. Finally, the new representations are passed to classification module to compute the final prediction for a steganalysis task. The simple architecture of
FCEM leads to fewer parameters, less computation and easier parallelization, making it a very excellent model to be widely applied in online services.

 \textbf{ The contribution of this work is:} \textbf{1)} we firstly proposed a method  solely based on attention mechanism to extract correlation features in steganalysis of VoIP stream  or even audio steganalysis task.
 \textbf{2)} experiments show that the proposed steganalysis method achieves excellent performance in detecting low embedding rates and short samples. What's more, proposed model are significantly reduced inference time when sample length is as short as 0.1s, making it can be easily utilized in online services.
% \textbf{We summarize our main contribution as follow:}
% \begin{itemize}
% \item[1)] To the best of our knowledge, we are the first to design a nerual network solely based on attention mechanism in steganalysis of VoIP stream  or even  audio steganalysis task. Experiment results verify the practicability of this mechanism and indicate that attention is a powerful alternative to extract correlation feature for steganalysis methods to solve similar problems.
% \item[2)] Experiments show that the proposed steganalysis method achieves excellent performance in detecting low embedding rates and short samples, which are the hardest part in VoIP steganalysis. What's more, the training and inference time of the proposed model are significantly reduced when sample length is as short as 0.1s, making it can be easily utilized in online services.
% \end{itemize}

The rest of the paper is structured as follows. In part 2,
we introduce the proposed method in detail. Part 3 shows experiments results. Finally, we make the conclusion and offer some aspirations for future research in part 4.
\section{THE PROPOSED NETWORK}
\begin{figure}[ht]
\centering
\includegraphics[width=3.8in]{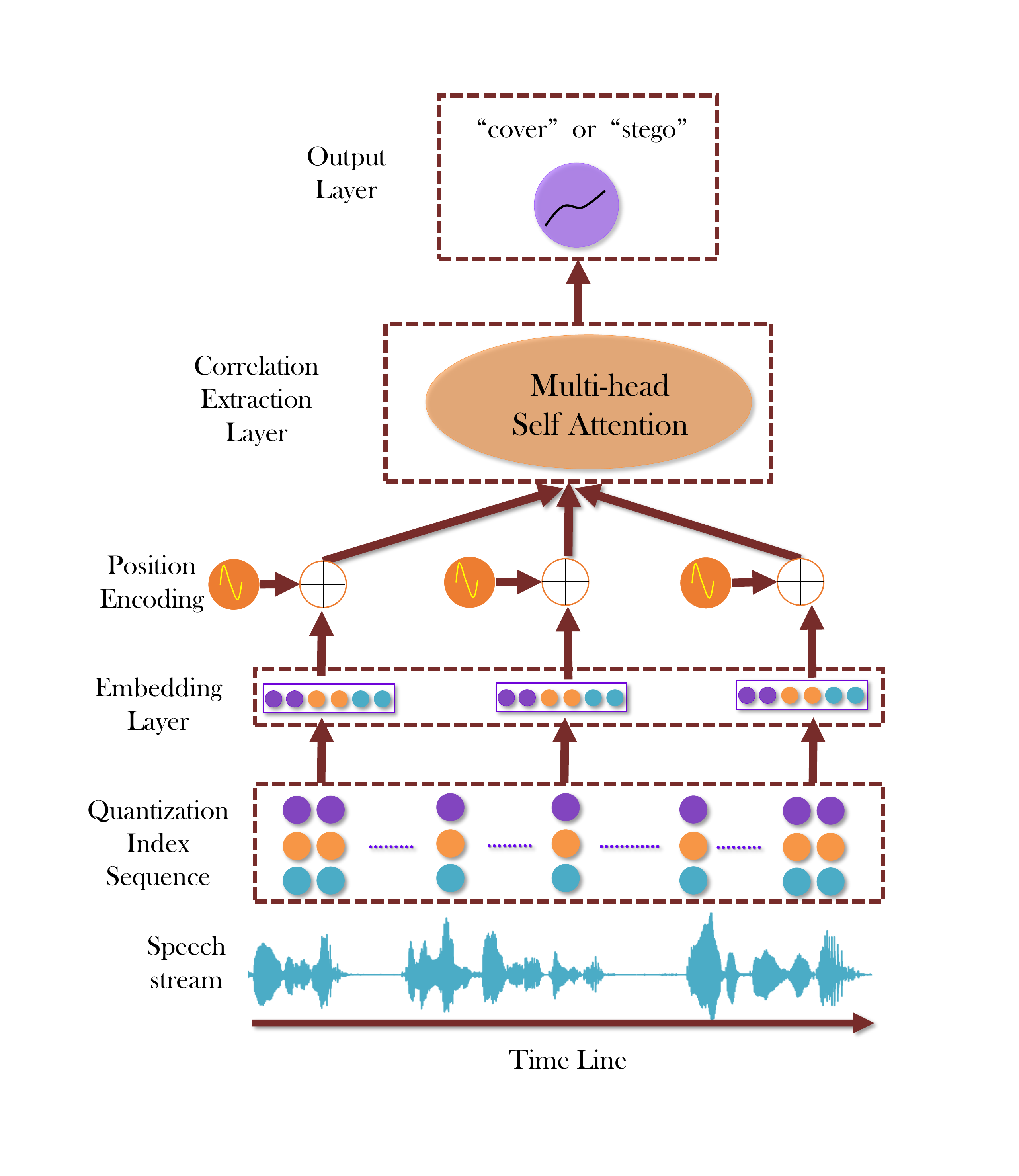}
\caption{Structure of Proposed Fast Correlation Extraction Model}
\label{structure}
\end{figure}

The architecture of the proposed Fast Correlation Extract Model model is illustrated in
Figure \ref{structure}. Each layer in our model from bottom
to top is explain in detail in the following parts.
\subsection{Input Layer-Quantization Index Sequence}
In general, low bit-rate speech codecs such as ITU-G.723.1 are widely used in VoIP communications for compressing the speech or audio signal components of streaming media. 
Most of the speech codecs are based on the linear predictive coding (LPC) \cite{LPC} model, which uses an LPC filter to analyze and synthesize acoustic signals between the
encoding and decoding endpoints. In LPC encoding, the LPC coefficients are first converted into Line Spectrum Frequency (LSF) coefficients. And
the LSFs are encoded by Vector Quantization (VQ). Quantization Index Sequence (QIS) is generated in the process of VQ, which can be denoted as follows:
\begin{equation}
\label{qismatrix}
QIS = \left[ {\begin{array}{*{20}{c}}
  {{s_{1,1}}}&{{s_{2,1}}} \\ 
  {{s_{1,2}}}&{{s_{2,2}}} \\ 
  {{s_{1,3}}}&{{s_{2,3}}} 
\end{array}\begin{array}{*{20}{c}}
  {.....\begin{array}{*{20}{c}}
  {} 
\end{array}\begin{array}{*{20}{c}}
  {{s_{T,1}}} \\ 
  {{s_{T,2}}} \\ 
  {{s_{T,3}}} 
\end{array}} 
\end{array}} \right],
\end{equation}
where $T$ is the total frame number in the sample window of the speech. In VoIP scenario, we can utilize sliding detection window  \cite{Huang2011Detection} to collect one or several continuous packets each time to construct these quantization index sequences for steganalysis .

%  In the process of VQ, an excitation
% signal is chosen through an analysis-by-synthesis search procedure
% in which the error between the original and reconstructed speech
% is minimized according to a perceptually weighted distortion. Open
% loop pitch search and closed loop pitch search make up the adaptive
% codebook search step of overlapping candidate vectors. After
% Algebraic codebook search,

QIM steganography divides quantization codebook of VQ into
several parts according to the embedded data. The VQ codewords in different position use independent  quantization codebooks, which will bring distortion to the characteristics of QIS. Thus, QIS can be taken as clues for steganalysis and we take it as model inputs.

\subsection{Embedding Layer}
Embedding Layer is a common approach to represent some objects using a low-dimension dense vector representation, which is a very popular technique in various task such as natural language processing \cite{devlin2018bert} and speech process \cite{yang2019fast}. Specifically, in our task, the weights of the Embedding layer are of the shape $(codeword\_size, embedding\_size)$. For each training sample, its input codewords are quantization integers in QIS. The integers are in the range of the $codeword\_size$. The Embedding layer transforms each quantization integer $i$ into the $i$-th line of the embedding weights matrix, i.e., 
\begin{equation}
{e_i} = {V}{s_i},
\end{equation}
where $V$ is an random initialized embedding matrix  for code-words, $s_i$ is corresponding one-hot vector created by quantization integer code-words in QIS and $e_i$ is an dense vector whose dimension is $embedding\_size$.
After the embedding layer,  representation of the $j$-th QIS frame can be denoted as:
\begin{equation}
{x_j} = [{e_{j,1}} \oplus {e_{j,2}} \oplus {e_{j,3}}],
\end{equation}
where $\oplus$ is the concatenation operator.

\subsection{Position Encoding}
The attention mechanism cannot distinguish between
different positions. So it is crucial to encode positions of
each code words. There are various ways to encode positions,
and the simplest one is to use an additional position
encoding module. In this paper, we try the approach
proposed in \cite{att}, which is formulated as
follows:
\begin{equation}
\begin{array}{l}
PE(t,2i) = \sin (t/{10000^{2i/d}})
\end{array}
\end{equation}
\begin{equation}
\begin{array}{l}
PE(t,2i + 1) = \cos (t/{10000^{2i/d}})
\end{array},
\end{equation}
where $t$ is the frame number, $i$ is the position in the embedded vector of $x_t$ and $d$ is the dimension of $x_t$. $d$ euqals to $3*embedding\_size$.
The position encodings are simply added to the input embeddings to encode position information.

\subsection{Correlation Extraction Layer}
Correlations between different frames are crucial in speech steganalysis \cite{Lin2018RNN}. Here, we adopt Multi-head self attention \cite{att} which has recently achieved
remarkable performance in modeling complicated relations between different code-words. Taking the $m$-th frame feature $x_m$ as an example, we will explain how to identify multiple meaningful correlation features involving feature $x_m$ based on such a mechanism. At first, We defined  the correlation between feature $x_m$ and feature $x_k$ under a specific attention head $h$ as follows:

\begin{equation}
\begin{array}{l}
\alpha _{m,k}^{(h)} = \frac{{\exp ({\varphi ^{(h)}}({x_m},{x_k}))}}{{\sum\nolimits_{l = 1}^M {\exp ({\varphi ^{(h)}}({x_m},{x_l}))} }}
\end{array}
\end{equation}
\begin{equation}
\begin{array}{l}
{\varphi ^{(h)}}({x_m},{x_k}){\rm{ = }}\left\langle {W_{query}^h{x_m},W_{key}^{(h)}{x_k}} \right\rangle 
\end{array},
\end{equation}
where ${\varphi ^{(h)}}\left( { \cdot , \cdot } \right)$ is an attention function which defines the correlation between feature $x_m$ and feature $x_k$. In most cases, it can be defined as inner product. $W_{query}^h,W_{key}^{(h)} \in {R^{d \times d'}} $ are transformation matrices which map the original embedding space ${R^{d}}$ into a new semantic space ${R^{d'}}$. Next we recalibrate representation of feature $x_m$ in subspace $h$ by combining all
relevant features guided by coefficients $\alpha _{m,k}^{(h)}$:
\begin{equation}
\hat{x}_m^{(h)} = \sum\nolimits_{k = 1}^M {\alpha _{m,k}^{(h)}(W_{value}^{(h)}{x_k})},
\end{equation}
where $W_{value}^{(h)} \in {R^{d \times d'}}$ and $M$ is the sequence length.
Since $\hat{x}_m^{(h)} \in {R^{d'}}$ is a combination of feature $x_m$ and its relevant
features under head $h$, it represents a new combinatorial feature
learned by our method. Furthermore, a feature is also likely to be
involved in different combinatorial features, and we achieve this by
using multiple heads, which create different subspaces and learn
distinct feature interactions separately. We collect combinatorial
features learned in all subspaces as follows:
\begin{equation}
\hat{x}_m = \hat x_m^{(1)} \oplus \hat x_m^{(2)} \oplus ... \oplus \hat x_m^{(H)},
\end{equation}
where $\oplus$ is the concatenation operator, and $H$ is the number of total
heads.

With such an interacting layer, the representation of each feature
$x_m$ will be updated into $\hat{x}_m$ to represent high-order correlation features.

% \begin{figure}
% \centering
% \includegraphics[width=4in]{matt.pdf}
% \caption{(left) Scaled Dot-Product Attention. (right) Multi-Head Attention consists of several
% attention layers running in parallel. In our model, $Q=K=V=x_m$}
% \label{matt}
% \end{figure}

\subsection{Output Layer}
The output of the correlation extraction layer is a set of feature vectors.  We simply concatenate them and then put them into a classification layer for final prediction. The process can be expressed as:
\begin{equation}
\hat{y} = \sigma (w \circ ({{\hat x}_1} \oplus {{\hat x}_2} \oplus ... \oplus {{\hat x}_T}) + b),
\end{equation}
where $\sigma (x) = \frac{1}{{1 + {e^{ - x}}}}$, $w$ and $b$ are parameter and bias term of linear transformation and $T$ is the total frame in the sample window. After this step, the model will output probabilities that samples belongs to cover or stego carrier. Besides, dropout is used to prevent over-fitting in this layer. 

The proposed model is trained in a supervised framework. Loss of training process is cross entropy loss. Parameters in the model can be update via minimizing the cross entropy loss based on gradient descent \cite{adam-ref}.

% \subsection{Losses}
% The proposed model is trained in a supervised framework. Loss of training process is cross entropy loss, which is defined as follows:
% \[L(y_i,\hat{y}_i) =  - \frac{1}{N}\sum\limits_{i = 1}^N {({y_i}\log ({\hat{y}_i}) + (1 - {y_i})\log (1 - {\hat{y}_i}))},\]
% where $y_i$ is the ground truth of sample type, $\hat{y}_i$ is the estimated label, $i$ indexes the training samples, and $N$ is the total number of training samples. Parameters in the model can be update via minimizing the cross entropy loss based on gradient descent \cite{adam-ref}.

\section{EXPERIMENT}
\subsection{Experimental Settings}
Our experiments were conducted in public dataset \footnote{https://github.com/fjxmlzn/RNN-SM}.The above dataset contained  different types of native speakers making it a good diversity in speech. Besides, each speech file in the
datasets was encoded according to the G.729a standard and secret data were embedded using CNV-QIM \cite{31-Xiao2008An} steganography with different embedding rates. In our experiment, in order to test the performance of proposed model in low embedding rates, we choose samples embeded secret data with embedding rates from $\%10$ to $\%50$ in step of $\%10$. Besides,  we cut samples into 0.1s, 0.3s, 0.5, 0.7s and 1s segments to test the model performance in short samples. The efficency test is also based on these samples. 
The hyperparameters in our model were selected via cross-validation on the trail set. More specifically, the $embedding\_size$ is 100. The head of attention $H=8$. The dimension of hidden units $d'$ in each head is 32. The dropout rate was 0.6 for the output layer. The batch size in training process was 256, and the maximal training epoch was set to 100. We used Adam \cite{adam-ref} as the optimizer for network training. Our model was implemented by Keras. We train all the networks on GeForce GTX 1080 GPU with 16G graphics memory. Prediction process is done both on previous GPU and on ”Intel(R) Xeon(R) CPU E5-2683 v3 \@2.00GHz”.

In order to validate the performance of our model, we compare with other state-of-the-art methods: QCCN \cite{27-Li2017Steganalysis}, RNN-SM \cite{Lin2018RNN}, R-CNN\cite{yang-cnnlstm}, and HRN \cite{yang2019hierarchical}. Metrics used in experiments are detection accuracy and inference time.
\subsection{Impact of Sample Length}
Durations of samples have significantly influences on detection accuracy of QIM steganography in VoIP streams. From the Table \ref{tab:duration}, we can see the accuracy of detecting short samples are harder than long samples. The reason is that longer sample will provide more clues for steganalysis. Besides, we can also observe that the proposed model can improve the performance 
in various length of sample especially sample as short as 0.1s which is more practical in actual application.  Thus, we can conclude that the proposed method can effectively detect the QIM steganography only by capturing a small segment speech stream of a monitored VoIP session.

% Table generated by Excel2LaTeX from sheet '检测性能'
\begin{table}[htbp]
  \centering
  \caption{Detection Accuracy of $20\%$ Embedding Rate under Short Samples}
  \setlength{\tabcolsep}{1mm}{
    \begin{tabular}{c|l|c|c|c|c|c}
    \hline
    \multicolumn{1}{c|}{\multirow{2}[4]{*}{\textbf{Language.}}} & \multicolumn{1}{c|}{\multirow{2}[4]{*}{\textbf{Method}}} & \multicolumn{5}{c}{\textbf{Sample Length (s)}} \\     &   & \textbf{0.1} & \textbf{0.3} & \textbf{0.5} & \textbf{0.7} & \textbf{1.0} \\
    \hline
    \multirow{5}[10]{*}{\textbf{Chinese}} & QCCN \cite{27-Li2017Steganalysis} & 54.17 & 58.15 & 61.19 & 62.33 & 65.67 \\
      & RNN-SM \cite{Lin2018RNN} & 58.03 & 67.29 & 72.01 & 75.08 & 77.49 \\
      & RCNN \cite{yang-cnnlstm} & 58.46 & 67.89 & 73.81 & 74.59 & 77.36 \\
      & HRN \cite{yang2019hierarchical} & 60.31 & 70.31 & 75.55 & 77.36 & 81.10 \\
      & \textbf{Ours} & \textbf{64.83} & \textbf{74.72} & \textbf{78.03} & \textbf{81.36} & \textbf{85.99} \\
    \hline
    \multirow{5}[10]{*}{\textbf{English}} & QCCN \cite{27-Li2017Steganalysis} & 57.29 & 58.10 & 61.49 & 63.48 & 66.06 \\
      & RNN-SM  \cite{Lin2018RNN} & 57.18 & 66.79 & 73.02 & 77.32 & 80.11 \\
      & RCNN \cite{yang-cnnlstm} & 58.67 & 68.06 & 75.06 & 76.94 & 79.01 \\
      & HRN  \cite{yang2019hierarchical} & 60.12 & 71.31 & 77.06 & 81.05 & 83.21 \\
      & \textbf{Ours} & \textbf{63.80} & \textbf{74.74} & \textbf{82.08} & \textbf{84.45} & \textbf{88.81} \\
    \hline
    \end{tabular}%
    }
  \label{tab:duration}%
\end{table}%

\subsection{Impact of Embedding Rates}
% Table generated by Excel2LaTeX from sheet '检测性能'
The embedding rate is an important factor influencing detecting accuracy. 
In order to prevent steganalysis, the two parties of communication often adopt low embedding strategy. Obviously, detection rates in low embedding rates make great sense in practical scenario. In this part, we mainly conduct steganalysis in low embedding rates to show the effectiveness of proposed model. From the Table \ref{tab:emb}, we can observe that the proposed model outperforms all of the previous methods in low embedding rates.
\begin{table}[htbp]
  \centering
  \caption{Detection Accuracy of 0.3s Samples under Low Embedding Rate}
  \setlength{\tabcolsep}{1mm}{
    \begin{tabular}{c|l|c|c|c|c|c}
    \hline
    \multicolumn{1}{c|}{\multirow{2}[4]{*}{\textbf{Language}}} & \multicolumn{1}{c|}{\multirow{2}[4]{*}{\textbf{Method}}} & \multicolumn{5}{c}{\textbf{Embedding Rates(\%)}} \\
     &   & \textbf{10} & \textbf{20} & \textbf{30} & \textbf{40} & \textbf{50} \\
    \hline
    \multirow{5}[10]{*}{\textbf{Chinese}} & QCCN \cite{27-Li2017Steganalysis} & 51.85 & 58.15 & 66.25 & 71.10 & 77.40 \\
     & RNN-SM \cite{Lin2018RNN} & 57.04 & 67.29 & 75.41 & 82.05 & 86.13 \\
     & R-CNN \cite{yang-cnnlstm} & 58.65 & 67.89 & 75.51 & 79.16 & 85.84 \\
     & HRN \cite{yang2019hierarchical} & 60.03 & 70.31 & 78.23 & 83.42 & 87.42 \\
     & \textbf{Ours} & \textbf{65.39} & \textbf{74.72} & \textbf{83.41} & \textbf{89.67} & \textbf{92.67} \\
    \hline
    \multirow{5}[10]{*}{\textbf{English}}
    & QCCN \cite{27-Li2017Steganalysis} & 52.30 & 58.10 & 63.05 & 72.90 & 76.30 \\
    & RNN-SM \cite{Lin2018RNN} & 57.18 & 66.42 & 76.74 & 81.00 & 88.25 \\
    & R-CNN \cite{yang-cnnlstm} & 58.78 & 68.06 & 76.56 & 82.70 & 87.07 \\
    & HRN \cite{yang2019hierarchical} & 61.20 & 70.16 & 79.65 & 85.35 & 89.20 \\
    & \textbf{Ours} & \textbf{65.53} & \textbf{74.74} & \textbf{84.25} & \textbf{89.48} & \textbf{92.95} \\
    \hline
    \end{tabular}%
    }
  \label{tab:emb}%
\end{table}%

\subsection{Evaluation of Time Efficiency of Proposed Model}
VoIP is a real-time services used in IP network which means that the detection of steganography in VoIP stream should conduct as fast as possible. Compared with previous machine learning framework, the proposed model utilizes attention have the advantage for easier parallelizable computation. Experiments in Table \ref{tab:inf} shows that our model drastically reduces the steganalysis time compared to recent works. Especially, we can see that even when the sample is as short as 0.1,  the proposed model accelerate the steganalysis speed more than twice.

% Table generated by Excel2LaTeX from sheet '检测性能'
\begin{table}[htbp]
  \centering
  
  \caption{Inference Performance of Different Models}
    \begin{tabular}{c|r|r|r|r|r}
    \hline
    \multicolumn{1}{c}{\multirow{2}[4]{*}{Method}} & \multicolumn{5}{|c}{Sample Length(ms)} \\
    & 100 & 300 & 500 & 700 & 1000 \\
    \hline
    RNN-SM \cite{Lin2018RNN} & 0.584 & 1.355 & 2.123 & 3.100 & 4.165 \\
    RCNN \cite{yang-cnnlstm} & 0.596 & 1.349 & 2.024 & 2.701 & 3.754 \\
    HRN \cite{yang2019hierarchical} & 0.268 & 0.349 & 0.398 & 0.421 & 0.512 \\
    \textbf{Ours} & \textbf{0.113} & \textbf{0.168} & \textbf{0.187} & \textbf{0.228} &\textbf{ 0.281} \\
    \hline
    \end{tabular}%
  \label{tab:inf}%
\end{table}%

\section{CONCLUSIONS}
VoIP is a very popular real time streaming media for steganography. Previous steganalysis methods always extracted correlation feature by adopting RNN, CNN and their combination in VoIP streams, which either has low computation effiency or is hard to model long dependency. In this paper, we introduced attention mechanisms to extract correlation in code-words for speech steganalysis tasks and designed a fast correlation extract network without any RNN/CNN structure to detect QIM based steganography in VoIP stream. Experiment results indicate that our model can outperform all of the previous work and achieve state-of-the-art results. Excellent detection performance of proposed method shows that steganalysis model based solely on attention is a powerful alternative to extract correlation feature for steganalysis of VoIP stream and can be easily extend to similar problems in audio steganalysis. 

% \section{REFERENCES}
% \label{sec:refs}
\bibliographystyle{IEEEbib}
\bibliography{strings}

\end{document}